\documentclass[11pt, showpacs]{revtex4}
\usepackage{graphicx, epsfig}

\def\xslash#1{{\rlap{$#1$}/}}
\def\half{\frac{1}{2}}
\def\beq{\begin{equation}}
\def\eeq{\end{equation}}
\def\beqa{\begin{eqnarray}}
\def\eeqa{\end{eqnarray}}
\def\iar{\begin{array}{l}}
\def\ear{\end{array}}

\begin{document}

\title{Pole wave-function renormalization prescription for unstable particles}
\author{Yong Zhou}
\affiliation{Beijing University of Posts and Telecommunications, School of Science, P.O. Box 123, Beijing 100876, China}

\begin{abstract}
We base a new wave-function renormalization prescription on the pole mass renormalization prescription, in which the Wave-function Renormalization Constant (WRC) is extracted by expanding the particle's propagator around its pole, rather than its physical mass point as convention. We find the difference between the new and the conventional WRC is gauge-parameter dependent for unstable particles beyond one-loop level, which will lead to some physical results gauge dependent under the conventional wave-function renormalization prescription beyond one-loop level.
\end{abstract}

%11.10.Gh Renormalization
\maketitle

\section{Introduction}

The conventional wave-function renormalization prescription extracts WRC by expanding the particle's propagator around its physical mass point in the LSZ reduction formula \cite{c1}. For scalar boson it is \cite{c2,c3,c5}
\beqa
  \frac{i}{p^2-m^2-\delta m^2+\Sigma(p^2)}&\sim&\frac{i}
  {(p^2-m^2)(1+Re\Sigma^{\prime}(m^2))+Re\Sigma(m^2)-\delta m^2+i\,Im\Sigma(p^2)} \nonumber \\
  &=&\frac{i\,(1+Re\Sigma^{\prime}(m^2))^{-1}}{p^2-m^2+i\,\epsilon}\,, \nonumber\\
  {\bf or}\hspace{3mm}&\sim&\frac{i}{(p^2-m^2)(1+\Sigma^{\prime}(m^2))
  +\Sigma(m^2)-\delta m^2}\,=\,\frac{i\,(1+\Sigma^{\prime}(m^2))^{-1}}{p^2-m^2+i\,\epsilon^{\prime}}\,,
\eeqa
where $\Sigma^{\prime}(m^2)\hspace{-1mm}=\hspace{-1mm}\partial\Sigma(m^2)/\partial p^2$, $\epsilon$ and $\epsilon^{\prime}$ are small quantities. But not long ago people propose that only the mass definition of the pole of the particle's propagator is gauge invariant \cite{c6} and physical results are only gauge invariant under the pole mass renormalization prescription \cite{c9}, so WRC must also be defined on the pole of the particle's propagator. Considering the fact that unstable particle's WRC must contain imaginary part \cite{c3,c5,cin,c4}, the new wave-function renormalization prescription for boson must be
\beq
  \frac{i}{p^2-m^2-\delta m^2+\Sigma(p^2)}\,\sim\,\frac{i}{(p^2-\bar{s})
  (1+\Sigma^{\prime}(\bar{s}))}\,=\,\frac{i\,(1+\Sigma^{\prime}(\bar{s}))^{-1}}{p^2-\bar{s}}\,,
\eeq
where $\bar{s}=m^2-i m\Gamma$ is the pole of the boson's propagator \cite{c6}. Note that the pole mass renormalization prescription has been used in Eq.(2).

For fermion the new wave-function renormalization prescription is a little complex. The fermion inverse propagator can be written as
\beq
  i S^{-1}(\xslash p)\,=\,\xslash p-m-\delta m+\Sigma(\xslash p)\,\equiv\,
  \xslash p(a\gamma_L+b\gamma_R)+c\gamma_L+d\gamma_R\,,
\eeq
where $\gamma_L$ and $\gamma_R$ are the left- and right- handed helicity operators, and the diagonal fermion self energy is
\beq
  \Sigma(\xslash p)\,=\,\xslash p\gamma_L\,\Sigma^L(p^2)+\xslash p\gamma_R\,\Sigma^R(p^2)
  +m(\gamma_L\Sigma^{S,L}(p^2)+\gamma_R\Sigma^{S,R}(p^2))\,.
\eeq
Expanding the fermion propagator around its pole we get \cite{cin,c4}
\beqa
  S(\xslash p)&=&\frac{i(\xslash p(a\gamma_L+b\gamma_R)-d\gamma_L-c\gamma_R)}
  {p^2 a b - c d} \nonumber \\
  &\sim&\frac{i(m+\delta m+\xslash p\gamma_L(1+\Sigma^L(\bar{s}))
  +\xslash p\gamma_R(1+\Sigma^R(\bar{s}))-m\gamma_L\Sigma^{S,R}(\bar{s})
  -m\gamma_R\Sigma^{S,L}(\bar{s}))}{(p^2-m^2+i\,m\,\Gamma)A}\,,
\eeqa
where $\bar{s}=m^2-i m\Gamma$ is the pole of the fermion propagator, and
\beqa
  A&=&(1+\Sigma^L(\bar{s}))(1+\Sigma^R(\bar{s}))+\bar{s}(
  \Sigma^{L\prime}(\bar{s})+\Sigma^{R\prime}(\bar{s})
  +\Sigma^{L\prime}(\bar{s})\Sigma^R(\bar{s})
  +\Sigma^{L}(\bar{s})\Sigma^{R\prime}(\bar{s}) ) \nonumber \\
  &+&m\Sigma^{S,L\prime}(\bar{s})(m+\delta m-m\Sigma^{S,R}(\bar{s}))
  +m\Sigma^{S,R\prime}(\bar{s})(m+\delta m-m\Sigma^{S,L}(\bar{s}))\,.
\eeqa

From Eq.(2) and Eq.(5) we can extract boson and fermion's WRC. In section 2 we will do this work. In section 3 we will evaluate the difference of unstable particle's WRC between the new and the conventional wave-function renormalization prescription and discuss the influence of the difference on physical results. Lastly we give our conclusion.

\section{Determination of wave-function renormalization constants}

In the LSZ reduction formula one needs to introduce two sets of WRC: the incoming WRC and the outgoing WRC \cite{c5,cin,c4}. For boson the incoming and outgoing WRC are introduced as follows \cite{c5}
\beq
  Z^{\half}\,=\,<\hspace{-1mm}\Omega|\phi(0)|\lambda\hspace{-1mm}>\,, \hspace{10mm}
  \bar{Z}^{\half}\,=\,<\hspace{-1mm}\lambda|\phi^{\dagger}(0)|\Omega\hspace{-1mm}>
  \,,
\eeq
where $\Omega$ is the interaction vacuum, $\phi$ is the boson's Heisenberg field, and $\lambda$ is the incoming or outgoing state of S-matrix element. According to the LSZ reduction formula we have from Eq.(2)
\beq
  Z^{\half}\bar{Z}^{\half}\,=\,(1+\Sigma^{\prime}(m^2-i\,m\,\Gamma))^{-1}\,.
\eeq
Another condition that boson's WRC must satisfy is \cite{c5}
\beq
  \bar{Z}\,=\,Z\,.
\eeq
Therefore we get
\beq
  \bar{Z}\,=\,Z\,=\,(1+\Sigma^{\prime}(m^2-i\,m\,\Gamma))^{-1}\,.
\eeq

For fermion the incoming and outgoing WRC are introduced as follows \cite{c5}:
\beq
  <\hspace{-1mm}\Omega|\psi(0)|\lambda\hspace{-1mm}>\,=\,Z^{\half}u\,, \hspace{10mm}
  <\hspace{-1mm}\lambda|\bar{\psi}(0)|\Omega\hspace{-1mm}>\,=\,\bar{u}\bar{Z}^{\half}\,,
\eeq
where $\psi$ is the fermion's Heisenberg field and
\beq
  Z^{\half}\,=\,Z^{L\half}\gamma_L+Z^{R\half}\gamma_R\,,\hspace{10mm}
  \bar{Z}^{\half}\,=\,\bar{Z}^{L\half}\gamma_R+\bar{Z}^{R\half}\gamma_{L}\,.
\eeq
The fermion propagator at resonant region can be expressed as \cite{c5,cin}
\beq
  S(\xslash p)\,\sim\,\frac{i\,Z^{\half}(\xslash p+m+i x)\bar{Z}^{\half}}{p^2-m^2+i\,m\,\Gamma}\,,
\eeq
where $x$ is a small quantity. Considering Eqs.(5,13) must be placed in the middle of on-shell spinors $\bar{u}$ and $u$ (or $\bar{\nu}$ and $\nu$) and the fact $\bar{u}\gamma_L u=\bar{u}\gamma_R u$ (or $\bar{\nu}\gamma_L \nu=\bar{\nu}\gamma_R \nu$), we obtain
\beqa
  Z^{L\half}\bar{Z}^{L\half}&=&(1+\Sigma^R(\bar{s}))/A\,, \nonumber \\
  Z^{R\half}\bar{Z}^{R\half}&=&(1+\Sigma^L(\bar{s}))/A\,, \nonumber \\
  Z^{L\half}\bar{Z}^{R\half}+Z^{R\half}\bar{Z}^{L\half}&=&
  (2m+2\delta m-m\Sigma^{S,L}(\bar{s})-m\Sigma^{S,R}(\bar{s}))/(A(m+i\,x))\,.
\eeqa
Another condition that fermion's WRC must satisfy is \cite{c5}
\beq
  \bar{Z}^L\,=\,Z^L\,, \hspace{10mm} \bar{Z}^R\,=\,Z^R\,.
\eeq
Therefore we get
\beqa
  \bar{Z}^L&=&Z^{L}\,=\,(1+\Sigma^R(\bar{s}))/A\,, \nonumber \\
  \bar{Z}^R&=&Z^{R}\,=\,(1+\Sigma^L(\bar{s}))/A\,.
\eeqa
Since the quantity $x$ is undefined in Eq.(13), the third equation of Eqs.(14) can be used to define $x$. At one-loop level we get \cite{cin}
\beq
  x\,=\,-\Gamma/2
\eeq
where $\Gamma$ is the fermion's decay width.

Now we have finished the definition of diagonal WRC. The off-diagonal WRC are out of our consideration, because they are different from the diagonal WRC under the meaning of the LSZ reduction formula.

\section{Gauge dependence of physical results under the conventional wave-function renormalization prescription}

Since unstable particle's WRC must contain imaginary part \cite{c3,c5,cin,c4}, the conventional wave-function renormalization prescription must be the second prescription of Eq.(1) for boson, i.e.  \cite{c5} (see Eq.(9))
\beq
  \bar{Z}_o\,=\,Z_o\,=\,(1+\Sigma^{\prime}(m^2))^{-1}\,,
\eeq
where the subscript $o$ represents the conventional wave-function renormalization prescription. Comparing with Eqs.(10) we find at two-loop level
\beq
  Z_o-Z\,=\,-i\,m\,\Gamma\,\Sigma^{\prime\prime}(m^2)\,.
\eeq
For unstable boson the difference is gauge-parameter dependent. For example for gauge boson W we obtain (see Fig.1)
\beqa
  &&Re[Z_{oW}-Z_W]_{\xi_W} \nonumber \\
  =&&\frac{\alpha^2}{288 s_w^2}\bigl{[} \sum_{i=e,\mu,\tau}
  (1-x_i)^2(2+x_i)+3\sum_{i=u,c}\sum_{j=d,s,b}|V_{ij}|^2\sqrt{1-2(x_i+x_j)+(x_i-x_j)^2} \nonumber \\
  \times&&(2-(x_i+x_j)-(x_i-x_j)^2) \bigr{]}(2\xi_W^3-3\xi_W^2-6\xi_W-5)\theta[1-\xi_W]\,,
\eeqa
where $Re$ takes the real part of the quantity, $\xi_W$ is the gauge parameter of W, the subscript $\xi_W$ denotes the $\xi_W$-dependent part of the quantity, $\alpha$ is the fine structure constant, $s_w$ is the sine of the weak mixing angle, $x_i=m_i^2/m_W^2$  and $x_j=m_j^2/m_W^2$ with $m_W$ the mass of W, $V_{ij}$ is the CKM matrix element \cite{c11}, and $\theta$ is the Heaviside function. Note that in the calculations we have used the program packages {\em FeynArts} and {\em FeynCalc} \cite{c10}.
\begin{figure}[htbp]
\begin{center}
  \epsfig{file=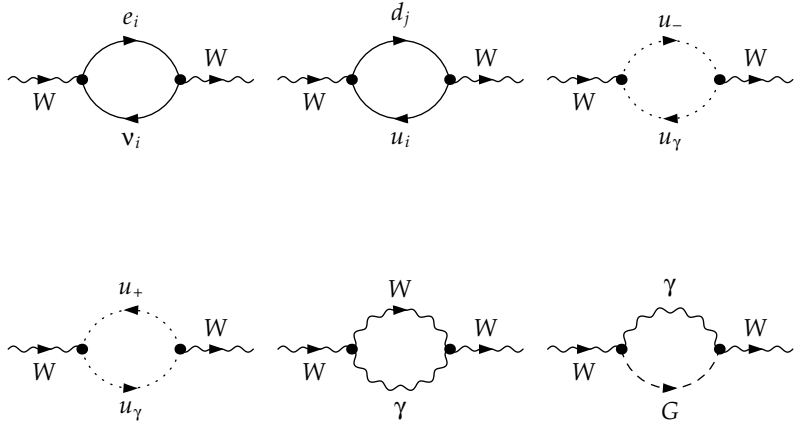, width=10.5cm} \\
  \caption{W one-loop self-energy diagrams containing imaginary part.}
\end{center}
\end{figure}

For fermion the conventional wave-function renormalization prescription must be \cite{c5}
\beqa
  \bar{Z}_o^L&=&Z_o^{L}\,=\,(1+\Sigma^R(m^2))/A_1\,, \nonumber \\
  \bar{Z}_o^R&=&Z_o^{R}\,=\,(1+\Sigma^L(m^2))/A_1\,,
\eeqa
where
\beqa
  A_1&=&(1+\Sigma^L(m^2))(1+\Sigma^R(m^2)) \nonumber \\
  &+&m^2(\Sigma^{L\prime}(m^2)
  +\Sigma^{R\prime}(m^2)+\Sigma^{L\prime}(m^2)\Sigma^R(m^2)
  +\Sigma^L(m^2)\Sigma^{R\prime}(m^2) ) \nonumber \\
  &+&m\Sigma^{S,L\prime}(m^2)(m+\delta m-m\Sigma^{S,R}(m^2)) \nonumber \\
  &+&m\Sigma^{S,R\prime}(m^2)(m+\delta m-m\Sigma^{S,L}(m^2))\,.
\eeqa
Comparing with Eqs.(16) we find
\beqa
  Z_o^L-Z^L&=&-i\,m\,\Gamma(2\Sigma^{L\prime}+\Sigma^{R\prime}
  +m^2(\Sigma^{L\prime\prime}+\Sigma^{R\prime\prime}+\Sigma^{S,L\prime\prime}
  +\Sigma^{S,R\prime\prime}))\,, \nonumber \\
  Z_o^R-Z^R&=&-i\,m\,\Gamma(\Sigma^{L\prime}+2\Sigma^{R\prime}
  +m^2(\Sigma^{L\prime\prime}+\Sigma^{R\prime\prime}+\Sigma^{S,L\prime\prime}
  +\Sigma^{S,R\prime\prime}))\,.
\eeqa
For unstable fermion the difference is also gauge-parameter dependent. For example for top quark we obtain (see Fig.2)
\beqa
  &&Re[Z_{ot}^L-Z_t^L]_{\xi_W} \nonumber \\
  =&&-\frac{\alpha^2}{128 s_w^4 x_t^3}
  \sum_{i=d,s,b}|V_{ti}|^2 B_i(x_t^2+(1-2 x_i)x_t+x_i^2+x_i-2) \nonumber \\
  \times&&\sum_{j=d,s,b}\frac{|V_{tj}|^2}{C_j}(x_t^3-x_j x_t^2-(3\xi_W^2+3 x_j\xi_W+2x_j^2)x_t \nonumber \\
  +&&2(\xi_W-x_j)^2(\xi_W+x_j))\theta[m_t-\sqrt{\xi_W}m_W-m_j]\,, \nonumber \\
  &&Re[Z_{ot}^R-Z_t^R]_{\xi_W} \nonumber \\
  =&&-\frac{\alpha^2}{256 s_w^4 x_t^3}
  \sum_{i=d,s,b}|V_{ti}|^2 B_i(x_t^2+(1-2 x_i)x_t+x_i^2+x_i-2) \nonumber \\
  \times&&\sum_{j=d,s,b}\frac{|V_{tj}|^2}{C_j}
  (x_t^3-(\xi_W+x_j)x_t^2-(\xi_W^2+4 x_j\xi_W+3 x_j^2)x_t \nonumber \\
  +&&(\xi_W-x_j)^2(\xi_W+3 x_j))\theta[m_t-\sqrt{\xi_W}m_W-m_j]\,,
\eeqa
where $m_t$ is the top quark's mass and $x_t=m_t^2/m_W^2$, and
\beq
  B_i\,=\,\sqrt{x_t^2-2(x_i+1)x_t+(x_i-1)^2}\,, \hspace{10mm}
  C_j\,=\,\sqrt{x_t^2-2(\xi_W+x_j)x_t+(\xi_W-x_j)^2}\,.
\eeq
\begin{figure}[htbp]
\begin{center}
  \epsfig{file=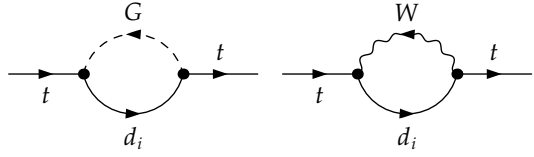, width=7cm} \\
  \caption{Top quark's one-loop self-energy diagrams containing imaginary part.}
\end{center}
\end{figure}

The gauge dependence of Eqs.(20,24) will lead to the decay widths
of some physical processes gauge-parameter dependent under the
conventional wave-function renormalization prescription.
Considering the physical process $W^{+}\rightarrow e^{+}\nu_e$,
since positive electron and electronic neutrino are stable
particles, their's WRC are same under the new and the conventional
wave-function renormalization prescription, therefore we only need
to consider the effect of $Z_{oW}$ on the gauge dependence of the
decay width. The result is shown in Fig.3 (the data is cited from
Ref.\cite{c12}).
\begin{figure}[htbp]
\begin{center}
  \epsfig{file=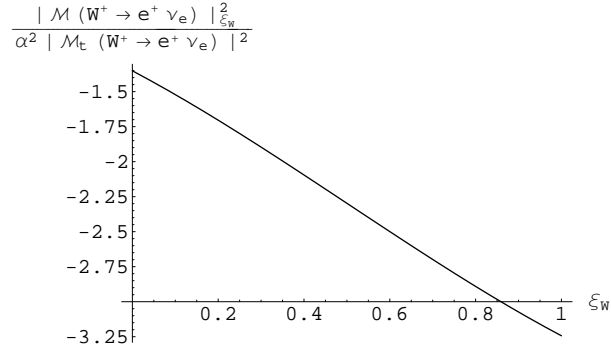, width=8cm} \\
  \caption{Gauge dependence of two-loop $|\cal{M}$$(W^{+}\rightarrow e^{+}\nu_e)|^2$ under
  the conventional wave-function renormalization prescription, where
  $\cal{M}$$_t(W^{+}\rightarrow e^{+}\nu_e)$ is the tree-level amplitude.}
\end{center}
\end{figure}
Considering another physical process $t\rightarrow W^{+}b$, since the decay width of bottom quark is zero at one-loop level, the bottom quark's WRC are same at two-loop level under the two wave-function renormalization prescriptions (see Eqs.(23)), so we only need to consider the effect of $Z_{ot}$ and $Z_{oW}$ on the gauge dependence of the decay width. The result is shown in Fig.4.
\begin{figure}[htbp]
\begin{center}
  \epsfig{file=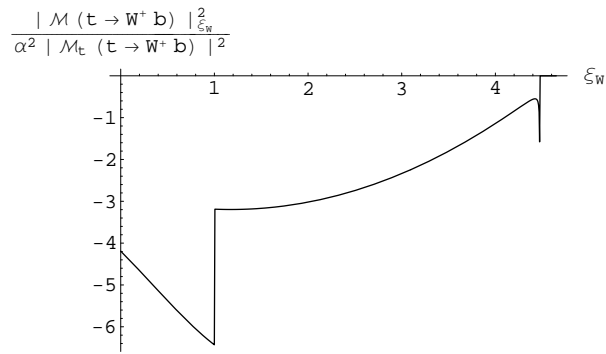, width=8cm} \\
  \caption{Gauge dependence of two-loop $|\cal{M}$$(t\rightarrow W^{+}b)|^2$ under
  the conventional wave-function renormalization prescription, where
  $\cal{M}$$_t(t\rightarrow W^{+}b)$ is the tree-level amplitude.}
\end{center}
\end{figure}
Fig.3 and Fig.4 show the gauge dependence of the two physical results under the conventional wave-function renormalization prescription is order of $O(1)$ at two-loop level, so the conventional wave-function renormalization prescription will affect the veracity of physical results beyond one-loop level.

\section{Conclusion}

The new wave-function renormalization prescription proposed here is based on the pole mass renormalization prescription, in which WRC is extracted by expanding the particle's propagator around its pole rather than its physical mass point as convention. The difference of the new WRC and the conventional WRC is gauge dependent for unstable particles beyond one-loop level. This will lead to some physical results gauge dependent under the conventional wave-function renormalization prescription beyond one-loop level.

\vspace{5mm} {\bf \Large Acknowledgments} \vspace{2mm}

The author thanks Prof. Cai-dian Lu for his devoted help.

\end{document}